\documentclass[aps,prl,twocolumn,groupedaddress,showpacs]{revtex4}
\usepackage{graphicx}
\begin{document}

\title{Resonant photon absorption and hole burning in Cr$_{7}$Ni antiferromagnetic rings}

\author{W. Wernsdorfer$^1$, D. Mailly$^2$, G. A. Timco$^3$, and R. E. P. Winpenny$^3$}

\affiliation{
$^1$Laboratoire Louis N\'eel, associ\'e \`a l'UJF, CNRS, BP 166, 38042 Grenoble Cedex 9, France\\
$^2$LPN, CNRS, Route de Nozay, 91460 Marcoussis, France\\
$^3$Department of Chemistry, The University of Manchester, Oxford Road, Manchester M13 9PL, UK}

\date{April 3, 2005}

\begin{abstract}
Presented are magnetization measurements on a crystal
of Cr$_{7}$Ni antiferromagnetic rings.
Irradiation with microwaves at frequencies between 1
and 10 GHz leads to observation of very narrow resonant photon absorption lines
which are mainly broadened by hyperfin interactions. 
A two-pulse hole burning technique allowed us to estimate the 
characteristic energy diffusion time.
\end{abstract}


\maketitle

Magnetic molecules are currently considered among the most
promising electron spin based quantum systems 
for the storing and processing of
quantum information~\cite{Awschalom02}. For this purpose,
ferromagnetic~\cite{Leuenberger01} and 
antiferromagnetic~\cite{Meier03a,Meier03b} systems 
have attracted an increasing interest~\cite{Carretta04,Troiani04a,Troiani04b}. 
In the latter case the quantum hardware is thought of as a collection
of coupled molecules, each corresponding to a different qubit. 
The main advantages would arise from the fact that 
they are extremely small and almost identical,
allowing to obtain, in a single measurement, statistical averages of
a larger number of qubits. The magnetic properties can
be modelled with an outstanding degree of accuracy.
And most importantly, the desired physical properties can be engineered
chemically.

Recently, the suitability of Cr-based antiferromagnetic molecular rings
for the qubit implementation has been proposed~\cite{Carretta04,Troiani04a,Troiani04b}.
The substitution of one metal ion in a Cr-based molecular 
ring with dominant antiferromagnetic
couplings allowed to engineer its level structure 
and ground-state degeneracy~\cite{Overgaard02,Larsen03}. 
A Cr$_{7}$Ni molecular ring was characterized by means
of low-temperature specific-heat and torque-magnetometry measurements, 
thus determining the microscopic parameters 
of the corresponding spin Hamiltonian. The
energy spectrum and the suppression of the 
leakage-inducing S-mixing render the Cr$_{7}$Ni molecule
a suitable candidate for the qubit implementation~\cite{Carretta05,Troiani04a,Troiani04b}.

In this paper we report the first micro-superconducting quantum interference device
(micro-SQUID)~\cite{WW_EPL04} studies of
the Cr$_{7}$Ni molecular ring.
Electron paramagnetic resonance (EPR) methods are
combined with high-sensitivity magnetization measurements. 
The magnetization detection could also be a Hall-probe 
magnetometer~\cite{Sorace03,Barco_PRL04,Bal04,Petukhov05,Bal05},
a standard SQUID~\cite{Cage05} or
a vibrating sample magnetometer~\cite{Candela65}.
We found very narrow resonant photon absorption lines
which are mainly broadened by hyperfin interactions. 
A two-pulse hole burning technique allowed us to estimate the 
characteristic energy diffusion time.

The Cr$_{7}$Ni molecular ring, is based on a homometallic ring with formula 
[Cr$_8$F$_8$(O$_2$CCMe$_3$)$_{16}$]. 
The eight chromium(III) ions lie
at the corners of a regular octagon~\cite{Overgaard02}.
Each edge of the octagon is bridged by one fluoride ion 
and two pivalate ligands. There is a large cavity 
at the centre of the ring. If a single chromium(III) ion 
is replaced by a metal(II) ion, for example nickel(II), 
this makes the ring anionic and a cation can be incorporated 
in the cavity. Thus we can make 
[H$_2$NMe$_2$][Cr$_7$NiF$_8$(O$_2$CCMe$_3$)$_{16}$]~\cite{Larsen03}.
If crystallized from a mixture of THF and MeCN 
the Cr$_{8}$ and Cr$_{7}$Ni compounds are isostructural, crystallizing 
in the tetragonal space group, $P$4.
 
The measurements were made in a dilution cryostat 
using a 20 $\mu$m sized single
crystal of Cr$_{7}$Ni. The magnetic probe was a micro-SQUID
array~\cite{WW_ACP01,WW_EPL04}
equipped with three coils allowing to apply a field in any
direction and with sweep rates up to 10 T/s.
The electromagnetic radiation was generated by a
frequency synthesizer (Anritsu MG3694A) triggered with a
nanosecond pulse generator.
This setup allows to vary continuously the frequency
from 0.1 Hz to 20 GHz, with pulse lengths form $\sim$1 ns to
continuous radiation~\cite{Thirion03}. Using a 50 $\mu$m sized 
gold radio frequency (RF) loop,
the RF radiation field was
directed in a plane perpendicular to the applied static field $\mu_0H$.
The microwave power of the generator could be varied from -80 to 20 dBm
(10$^{-11}$ to 10$^{-1}$ W). 
The sample absorbes only a small fraction of the generator power.
This fraction is however
proportional to the microwave power of the generator.

\begin{figure}
\begin{center}
\includegraphics[width=.45\textwidth]{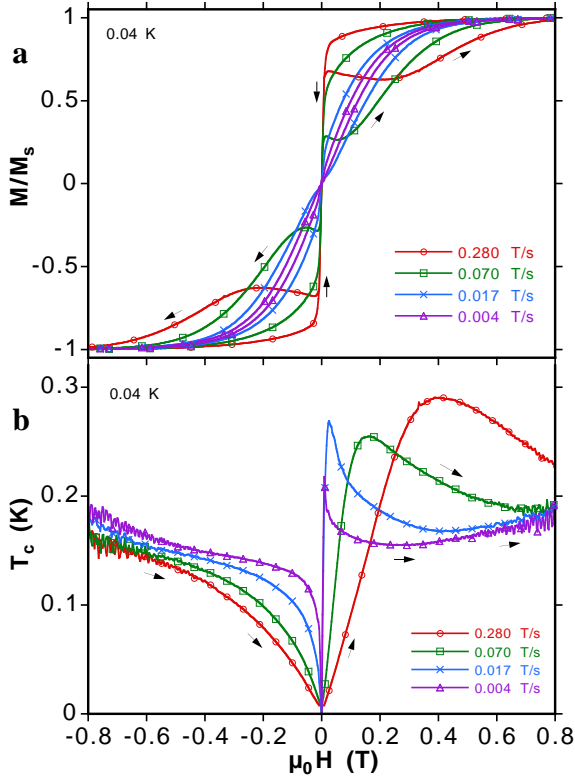}
\caption{(a) Magnetization ($M$) hysteresis loops for several field sweep 
rates at a cryostat temperature of 0.04 K. The loops are normalized
by the saturation magnetization $M_{\rm s}$ at 1.5 T. 
(b) Spin temperature $T_{S}$ 
for field sweeps from negative to positive fields, obtained 
by inversion of Eq. 1 where $M(T_{S})$ are the data in (a).}
\label{fig1}
\end{center}
\end{figure}

Fig. 1a shows magnetization versus applied field curves
for several field sweep 
rates at a cryostat temperature of 0.04 K.
The magnetization loops exibits a clear
hysteresis which is characteristic for the phonon-bottleneck regime 
with a spin-phonon relaxation time 
to the cryostat of a few seconds~\cite{ChiorescuV15PRL00}.
In order to quantify the out-of-equilibrium effect,
Fig. 1b presents the same data as in Fig. 1a but
the magnetization $M$ is converted into a spin temperature $T_{S}$
using the equation~\cite{Abragam70}:
\begin{equation}
    M(T_S)/M_{\rm s} = tanh(g\mu_{\rm B}S\mu_0H/k_{\rm B}T_S)
\label{eq_Ts}
\end{equation}
with $S = 1/2$ and  $g = 2.1$~\cite{Larsen03}. 
Fig. 1b shows clearly a strong adiabatic cooling when
sweeping the field down to zero field. Note that this cooling
mechanism might be used before qubit operations to
reach extremely low temperatures even at relatively high cryostat
temperatures. High frequency noise from the RF-loop around
the sample leads to spin temperatures at 1 T being higher than the 
cryostat temperature.

\begin{figure}
\begin{center}
\includegraphics[width=.45\textwidth]{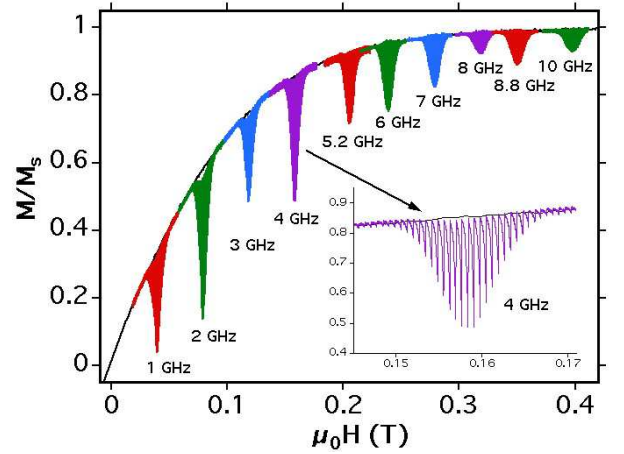}
\caption{Magnetization curves 
measured with and without irradiation. The cryostat temperature 
was 40 mK and the field sweep rate of 0.00014 T/s was
slowly in order to keep the the system at equilibrium. 
The electromagnetic radiation was pulsed with 
a period of 4 s and a pulse length of 1 $\mu$s. The RF frequencies
are indicated. Inset: Enlargement of the 4 GHz resonance. 
The fine structure is due to the RF pulses.}
\label{fig2}
\end{center}
\end{figure}

Fig.2 shows magnetization curves $M(H)$ 
in the quasi-static regime with a field 
sweep rate slow enough (0.00014 T/s) to keep the system 
at equilibrium. During the field sweep, RF pulses
were applied to the sample with a pulse length of 1 $\mu$s
and a period of 4 s between each pulse.
Depending on the RF frequency, clear dips are observed
which result from resonant
absorptions of photons associated with spin
transitions between the quantum numbers $m_s = 1/2$ and $-1/2$.
After each pulse, the magnetization relaxes back to 
the equilibrium magnetization (see the fine structure 
in the inset of Fig. 2).

\begin{figure}
\begin{center}
\includegraphics[width=.45\textwidth]{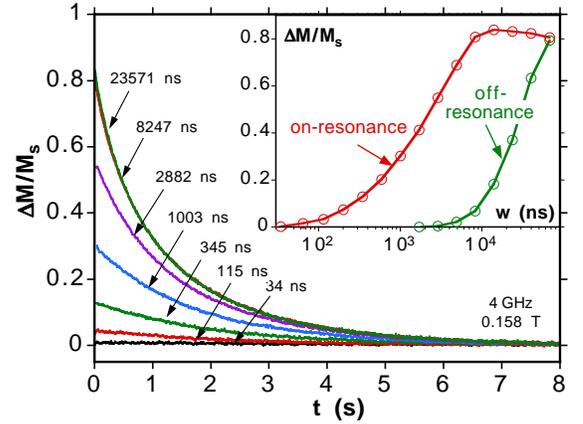}
\caption{Relaxation of magnetization after RF pulses
of 4 GHz were applied. The pulse lengths $w$ are indicated. 
Inset: magnetization variation $\Delta M$ 
after a RF pulse versus the pulse length $w$ for a
on-resonance field (0.1582 T) and off-resonance field (0.1722 T).}
\label{fig3}
\end{center}
\end{figure}

Typical relaxation measurements at a constant applied 
field after RF pulses of different durations are shown in Fig. 3. 
The relaxation is exponential with the rate 
independent of the pulse length. Detailed studies
showed that the relaxation rate is dominated by the
phonon-bottleneck regime, that is the spin-phonon relaxation time 
to the cryostat.

The inset of Fig. 3 presents the change of magnetization $\Delta M$ 
between the magnetization before and
after the pulse as a function of the pulse length $w$. $\Delta M$ 
increases linearly with $w$ for short pulses of few tens of ns. It 
saturates for $w \approx$ 10 $\mu$s and decrease for very long pulses 
because of cryostat heating effects. Non-resonant photon absorption is 
also observed for very long pulses.

The resonant photon absorption lines are often taken to estimate
a lower bound on the decoherence time of a qubit. We therefore
investigated in more detail the line width observed in Fig. 2. 
Fig. 4a presents a 
typical power dependence of the line width for continuous irradiation
at 4.2 GHz. Resonant photon absorption is clear visible for
a generator power larger than -60 dBm (1 nW). The line saturated at
about -20 dBm (10 $\mu$W). Fig. 4b presents the absorption line 
for the pulsed technique (see Fig. 2) for several pulse lengths
and a generator power of 15 dBm (32 mW).
The resonant photon absorption is clearly visible for pulse lengths
longer than 10 ns. Note that the pulse widths in Fig. 4a are
nearly two times larger than those in Fig. 4b.

\begin{figure}
\begin{center}
\includegraphics[width=.45\textwidth]{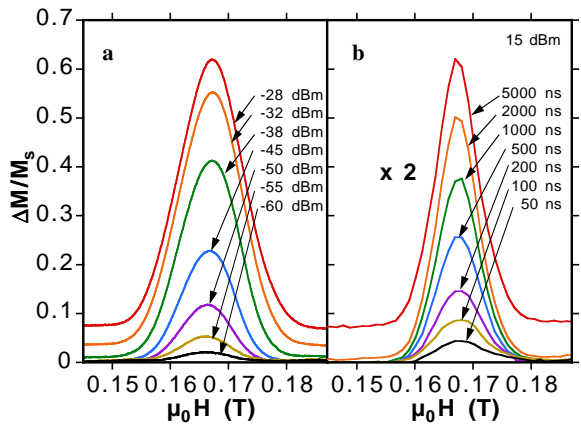}
\caption{(a) Magnetization variation $\Delta M$ 
between the equlibrium curves measured without 
and with continuous irradiation.
The microwave frequency was 4.2 GHz. 
The microwave power of the generator are indicated.
(b) Magnetization variation $\Delta M$ after a RF pulse
of 4.2 GHz and several pulse lengths.
The cryostat temperature was 40 mK. $\Delta M$ is multiplied
be a factor two.}
\label{fig4}
\end{center}
\end{figure}

\begin{figure}
\begin{center}
\includegraphics[width=.45\textwidth]{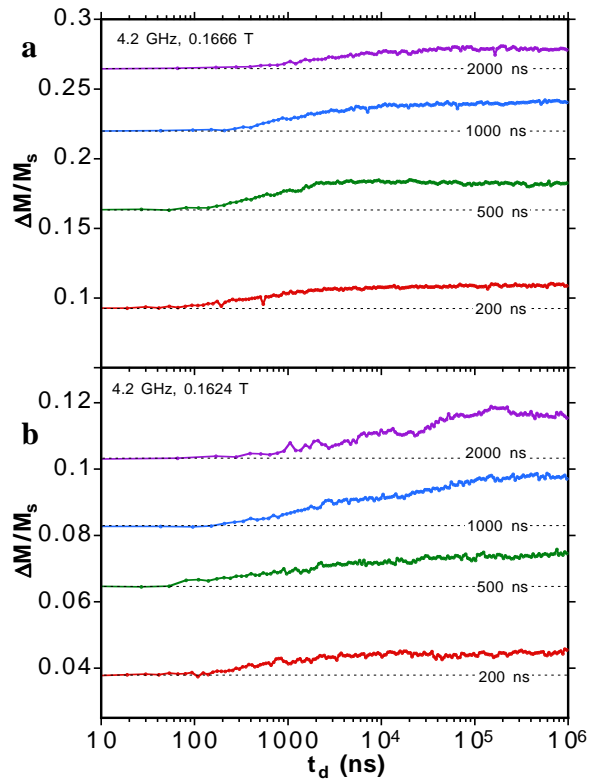}
\caption{Magnetization variation $\Delta M$ after two pulses 
of equal length as a function of delay time $t_{\rm d}$ between the pulses,  
(a) at an applied field of of 0.1666 T corresponding to
maximum resonance absorption, and (b) at 0.1624 T 
The RF frequency is 4.2 GHz and the  pulse lengths are indicated.
The dotted lines indicate $\Delta M$ for $t_{\rm d} = 0$. }
\label{fig5}
\end{center}
\end{figure}

In order to shed light on the origin of the line width broadening,
we developed a two pulse hole burning technique. The first pulse excites 
a fraction of spins that are in resonance during the pulse.
In case of an inhomogeneous broadened line, the first pulse 
burns a hole into the line. Then, after a delay time $t_{\rm d}$
a second pulse of the same frequency is applied. 
For $t_{\rm d} = 0$ the two pulses are equivalent to
one long pulse that excits a certain amount of spins (dotted
lines in Fig. 5).
However, for non-zero delay times $t_{\rm d}$ the second pulse
probes the evolution of spin excitation in the sample. For an
inhomogneous broadened line, the second pulse probes whether
the  burned hole of the first pulse evolved during the delay time.
Fig. 5 shows the resulting magnetization variation $\Delta M$ 
after two pulses as a function of delay time. It is shown that 
for $t_{\rm d} > $ 100 ns, $\Delta M$ is clearly larger than
for $t_{\rm d} = 0$. This result suggests that the absorption
lines in Figs. 2 and 4 are inhomogeneously broadened (at these time
scales). The first pulse burns a hole into the line which
starts to fill during the delay time. The filling time is clearly
faster for an applied field in the center of the line (Fig. 5a) 
than at the border (Fig. 5b). It depends also on the pulse length:
the longer the pulse length, the later starts the filling of the hole.
Because the magnetization is relaxing back to equilibrium
at a time scale much longer (phonon bottleneck regime),
the hole filling can only be due to spins that are close
to the resonance condition. Due to spin-spin interactions,
the excited spin can give their energy to those spins, that is
the energy diffuses from the excited spins to spins that are
close to the resonance condition. The hole
filling time is therefore dominated by spin-spin interactions
and it can be called a energy diffusion time.

In our case of an assembly of identical spins, the line broadening
is mainly due to dipolar and hyperfine interactions. 
The dipolar coupling energy can be estimated
with $E_{\rm dip}/k_{\rm B} \approx (g\mu_{\rm B}S)^2/V \approx 0.1$ mK
(S= 1/2 and $V$ = 6.3 nm$^3$)~\cite{Troiani04b}.
The hyperfine coupling with the nuclear spins 
can be obtained by considering the dipolar interaction 
of one Cr ion ($s = 3/2$) with the neighboring 
F nucleus having a nuclear spin $I$ = 1/2. 
With $g_{\rm F} = +5.26$ and the distance 
of $d$ = 0.2 nm between F and Cr ions, 
the interaction energy is about 0.4 mK
for each of the eight F nuclear spins~\cite{Troiani04b}. 
The hyperfine line broadening of all eight F nuclear spins
is about 3 mK which corresponds to 5 mT, in good
agreement with the observed line widths in Figs. 2 and 4.
These line widths give a decoherence time of about 3 ns~\cite{Troiani04b}.
Substantial reduction of the hyperfine broadening 
might be achieved by 
substituting the F ions with OH groups.
The dipolar coupling can also be reduced by
doping the crystal of Cr$_7$Ni molecules
with Cr$_8$ molecules.

In conclusion, we presented magnetization measurements on a crystal
of Cr$_{7}$Ni antiferromagnetic rings.
Irradiation with microwaves at frequencies between 1
and 10 GHz leads to observation of very narrow resonant photon absorption lines
which are broadened by spin-spin interactions. 
A two-pulse hole burning technique allowed us to estimate the 
characteristic energy diffusion time.

This work was supported by the EC-TMR Network 
ÒQuEMolNaÓ (MRTN-CT-2003-504880), EPSRC(UK), INTAS, 
CNRS and Rhone-Alpe funding.


\end{document}